# Enhancing Adjoint Optimization-based Photonics Inverse Design with Explainable Machine Learning

*Christopher Yeung[1,4], David Ho[1], Benjamin Pham[1], Katherine T. Fountaine[4], Zihan Zhang[2,3], Kara Levy[1], and Aaswath P. Raman[1,*]*

[1]Department of Materials Science and Engineering, University of California, Los Angeles, CA 90095, USA
[2]Department of Mathematics, University of California, Los Angeles, CA 90095, USA
[3]Department of Statistics, University of California, Los Angeles, CA 90095, USA
[4]NG Next, Northrop Grumman Corporation, Redondo Beach, CA 90278, USA
*Corresponding Author: <u>aaswath@ucla.edu</u>

**ABSTRACT:** A fundamental challenge in the design of photonic devices, and electromagnetic structures more generally, is the optimization of their overall architecture to achieve a desired response. To this end, topology or shape optimizers based on the adjoint variables method have been widely adopted due to their high computational efficiency and ability to create complex freeform geometries. However, the functional understanding of such freeform structures remains a black box. Moreover, unless a design space of high-performance devices is known in advance, such gradient-based optimizers can get trapped in local minima valleys or saddle points, which limits performance achievable through this inverse design process. To elucidate the relationships between device performance and nanoscale structuring while mitigating the effects of local minima trapping, we present an inverse design framework that combines adjoint optimization, automated machine learning (AutoML), and explainable artificial intelligence (XAI). Integrated with a numerical electromagnetics simulation method, our framework reveals structural contributions towards a figure-of-merit (FOM) of interest. Through an explanation-based reoptimization process, this information is then leveraged to minimize the FOM further than that obtained through adjoint optimization alone, thus overcoming the optimization's local minima. We demonstrate our framework in the context of waveguide design and achieve between 39% and 74% increases in device performance relative to state-of-the-art adjoint optimization-based inverse design across a range of telecom wavelengths. Results of this work therefore highlight machine learning strategies that can substantially extend and enhance the capabilities of a conventional, optimization-based inverse design algorithm while revealing deeper insights into the algorithm's designs.

# Introduction

Effectively optimizing nanophotonic structures is key to their use in a broad range of optical applications. For example, photonic integrated circuits, metasurfaces, and guided-wave systems can be geometrically manipulated at subwavelength scales to deliver a wide range of functionalities[1-5]. However, a large design space must be rapidly explored in order to optimize the geometry for a particular application. To effectively navigate such a design space, gradient-based optimization algorithms such as the adjoint variables method have been widely adopted to design non-intuitive or irregularly-shaped electromagnetic structures that are highly efficient at accomplishing a particular goal. By calculating the shape derivatives at all points in space using only two electromagnetic simulations per iteration[6], adjoint optimizations are orders of magnitude more computationally efficient than alternative optimization methods and capable of achieving state-of-the-art performance[6-9].

Although adjoint optimization-based methods have been successfully applied to a variety of photonic systems[10-13], the method's reliance on gradient-based information means that the method is local in nature, and therefore bounded by the corresponding limitations. Specifically, since the design space for electromagnetic structures is predominantly non-convex, adjoint optimizations (or gradient-based optimization algorithms in general) are susceptible to getting stuck in local minima valleys or saddle points (hereon collectively referred to as local minima)[14,35]. Thus, unless a region of high-performance devices is known in advance, multiple optimization runs are needed (typically by using random starting points) to arrive at a single optimization target[15]. To overcome these limitations, recent efforts have combined machine learning (ML) with adjoint optimization. For example, population-based inverse design was demonstrated using global topology-optimization networks, or GLOnets[16], which integrate the adjoint method directly into the training process. Alternative strategies also include the integration of ML and adjoint optimization as a two-step process, where the ML-component performs an initial global-search approximation, then the optimization improves design performance further[16-18,41]. Although both approaches can improve upon the algorithm's performance, the underlying issue of local minima trapping remains unaddressed, since the integration and use of a gradient-based optimizer inherently indicates that the issue is still present. In this regard, metaheuristic techniques such as simulated annealing have been proposed to escape local minima in the search process[19,20]. This method may directly address the issue of local minima trapping through neighbor-based exploration, but its applications in photonics shape and topology optimization have been severely limited due to relatively low computational efficiency (on the order of 1,000 iterations)[21,22].

To comprehensively address the issue of local minima trapping, we seek to identify the root of the problem and ask: what caused the algorithm to get trapped in the first place? In the context of the optical structures being optimized, arriving at certain geometric elements and their resulting electromagnetic response must be responsible for (or contribute to) guiding the optimization towards suboptimal results. To discover the geometric features responsible for local minima trapping, and to then overcome them, we employ an explainable artificial intelligence (XAI) based approach. XAI serves as a promising candidate for addressing local minima trapping due to its well-known ability to reveal a model's decision making process as well as the contributing factors thereof (*i.e.*, addressing the black box problem)[23-25]. For example, XAI can reveal the spatial regions of a nanophotonic structure that contribute to the presence or lack of an absorption peak[26]. Thus, to explain the causes of local minima trapping in gradient-based adjoint optimization, and subsequently use this information to prevent the optimization from converging onto suboptimal states, we present an XAI-based framework that utilizes the relationships between device efficiency and nanoscale structuring to increase optimization performance.

## Methods

We demonstrate our optimization framework in the context of Y-splitter waveguide design (Figure 1a), where the objective is to optimize the shape of the silicon-oxide interface to maximize power transmission efficiency from an input port to two output ports of the same width. We represent the objective function as a decreasing figure-of-merit (FOM) which ranges from 1 to 0, where 0 represents ideal performance. With this definition in place, adjoint shape optimizations are performed on an initial Y-splitter design to minimize the FOM (Figure 1b) across a range of target wavelengths in the telecommunications range (1.3 to 1.8 µm). The design and FOM information from the optimizations are used in conjunction with neural architecture search to automate the training of an ML model (AutoML). The model learns the relationships between device structure and performance by accurately predicting the FOM and target wavelength of an input design (Figure 1c). We then use a suite of XAI algorithms, SHapley Additive exPlanations or SHAP (a post hoc explanation technique based on game theory[27]), on the model to extract the structure-performance relationships as "feature explanation" heatmaps (Figure 1d). By interrogating our trained ML model, the explanations here inform the structural features that contribute to the FOM of interest. Using this information, we devise a boundary extraction algorithm that takes the explanations and makes guided design changes (Figure 1e). These design changes then provide a new starting point for the local adjoint optimization method, which allows the method to reach lower FOMs than before (at multiple target wavelengths).

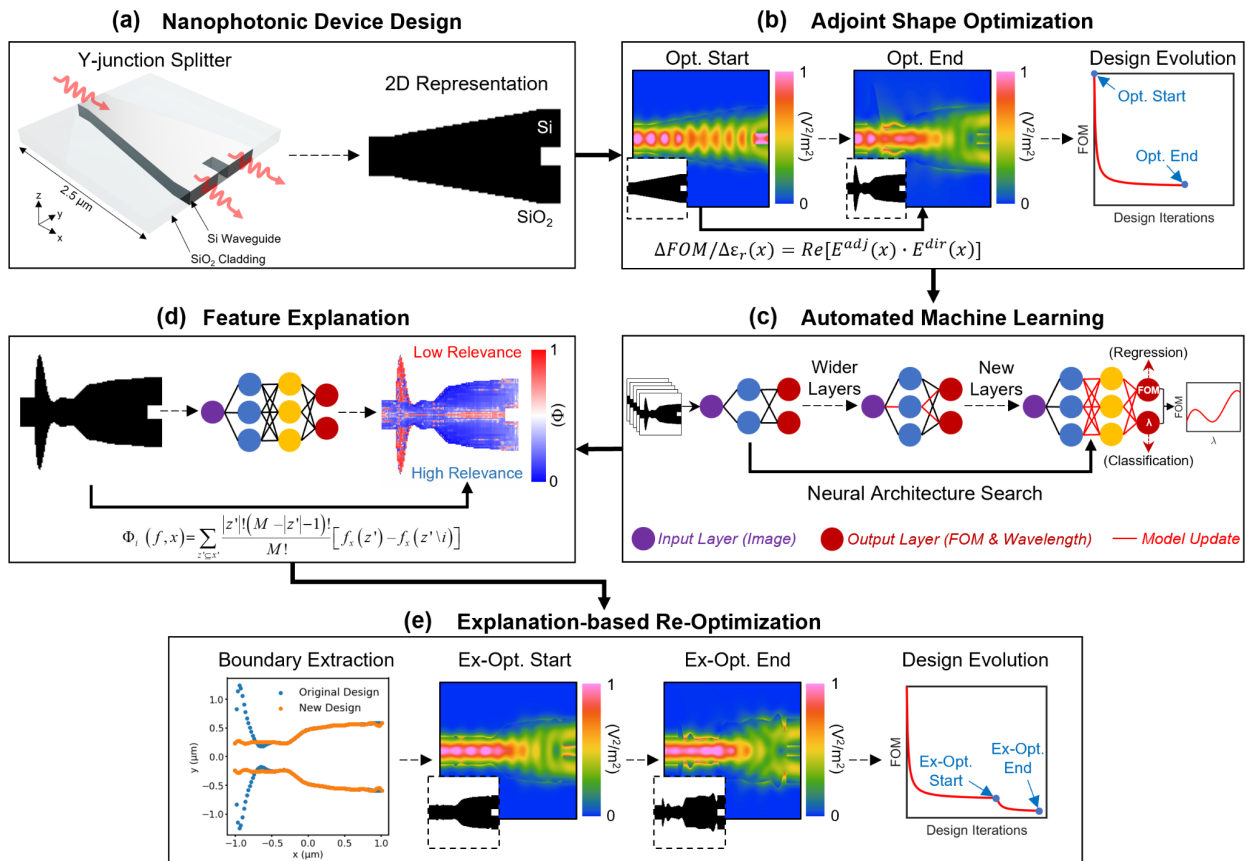

**Figure 1.** (a) Nanophotonic device optimization: silicon-on-insulator Y-junction splitter for telecom applications. (b) Multiple adjoint optimization runs are applied to the Y-splitter design at various target wavelengths. (c) Results of the optimizations are used as training data in automated machine learning (AutoML) to train a neural network, where the inputs are images and the outputs are device figure-of-merits (FOM) and target wavelengths. (d) Explainable AI algorithms are used on the neural network to capture feature explanations, (e) which are used to optimize device performance further by allowing the algorithm to escape its local minima.

# Results and Discussion

**Adjoint Optimization and Convolutional Neural Network Training**

We first developed our training dataset by performing multiple adjoint shape optimization runs on a starting Y-splitter design (Figure 2a). We applied a widely-adopted implementation of the adjoint method that is integrated with a commercial finite-difference time-domain (FDTD) solver[28]. The 2D cross-sections of the Y-splitter designs are represented as black and white images, where the black and white pixels represent the permittivity of silicon and $SiO_2$, respectively. In our configuration of the adjoint method, the optimizable region is the area between the input and output ports, while the port sizes remain fixed. The optimizable geometry within this region is defined using the level set method and cubic spline interpolations[28,29]. Each optimization run was performed on randomized starting designs (waveguide structures with 25%, 35%, 40%, and 50% fill fractions; collectively shown in Figure S2) using different operating wavelengths as optimization objectives (1.3 to 1.8 μm in 0.1 μm steps) to produce a collection of device designs with gradual performance improvements. Performance improvement is indicated by a decreasing FOM (as design iteration increases), until a plateau is reached. As shown in Figure 2a (here, the 35% fill fraction starting design), each design iteration consists of a forward and adjoint (*i.e.*, time-reversed) simulation, which calculates the shape derivative over the entire optimizable region and modifies the geometry (per iteration) in proportion to the FOM gradient[6]. At the final iteration "N" (which may vary across each optimization run), device geometry is tailored to achieve maximum attainable performance with respect to the sought target. In our application of the adjoint method, the FOM represents the power coupling of guided modes, and is defined as:

$$FOM = \frac{1}{P(\lambda)} \int |T_0(\lambda) - T(\lambda)| \, d\lambda, \qquad (1)$$

where $\lambda$ is the evaluated wavelength, $T$ is the actual power transmission through the output ports, $T_0$ is the ideal power transmission, and $P$ is the source power (in Watts). Thus, the FOM is the difference between the input and output transmission normalized by the power injected by the source, which results in maximum performance at 0. Figure 2b shows the results of each optimization run, where the collective FOM information and corresponding designs (at each iteration; not including the starting design) are used as training data for deep learning.

From the optimized structures shown in Figure 2, unique geometries are obtained for each target wavelength, which in turn yield a range of FOM values. Therefore, the FOM and target wavelength are coupled with one another, and both are dependent on the waveguide structure. Thus, to ensure that our model simultaneously learns both of these properties, which in turn captures more information regarding the structure than models trained on the properties individually, we designed a single neural network that takes the Y-splitter geometries as inputs (here, 128×64 pixel images, or 2.5×1.25 µm$^2$ domains) and outputs both FOM and target wavelength. In the particular design space we explored, over 600 input and output pairs were generated for the neural network. For ease of training, target wavelengths were converted into categorical labels, where a position-specific output node value of 1 represents the wavelength of a specific design, while the other positional nodes equal 0. For example, a target wavelength ($T_\lambda$) of 1.3 µm is represented as $T_{1.3}$ = [*1,0,0,0,0,0*], 1.4 µm is $T_{1.4}$ = [*0,1,0,0,0,0*], and this pattern is repeated up to 1.8 µm. Alternatively, *argmax($T_\lambda$)*, or the index of the maximum value along the vector, represents the target wavelength. Combined with a floating point value (ranging from 0 to 1) to serve as the FOM, we devise a training data structure that is amenable to both classification and regression-based tasks. With this input-output relationship defined, as well as a 90:10 training-validation data split, we use neural architecture search (AutoKeras[30]) with image blocks to automate the deep learning process by testing different model variants across multiple trials. We observe that the optimal architecture was identified after 12 trials, which had a validation loss of 9.1×10$^{-5}$. The final training and validation losses of each trial are presented in Figure S1a, and the evolution of the convolutional neural network (CNN) architecture from the first trial to the last is shown in Figure S1b. The optimized CNN possesses five convolutional blocks (with 512, 256, 128, 64, and 32 filters, respectively) followed by a dense layer. Each block contains Leaky ReLu, batch normalization, and max pooling layers. Training progression of the optimized architecture is shown in Figure 2c, where a strong convergence between the training and validation losses can be observed. We further verified our model's performance through cross validation and overfitting analyses (found in Table S1, S2 and Figures S12, S13 of the Supporting Information).

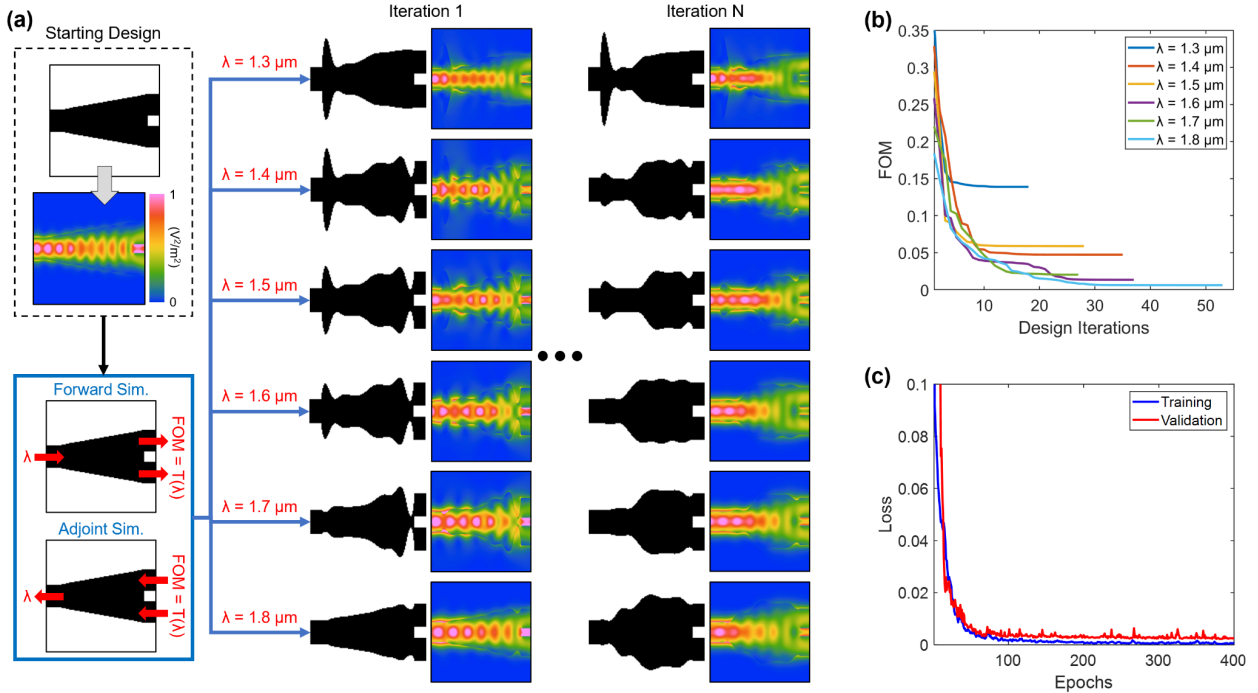

**Figure 2.** (a) Training data generation. Adjoint optimization runs are performed on random starting designs (35% fill fraction starting design shown) across target wavelengths ranging from 1.3 μm to 1.8 μm to produce high-performance devices in the telecom window. (b) FOM vs. design iterations across each optimization run. (c) Training and validation losses for the AutoML-optimized neural network shows high training accuracy and model convergence.

**Structure Explanation and Re-Optimization**

After training our machine learning model, we next sought to explain the relationship between the overall shape and FOM such that this information can be leveraged to potentially further optimize the devices, and overcome any local minima the adjoint method may have arrived at. To verify that the model properly learned the structure-FOM relationship, we passed the final design iterations (of each target wavelength) into the trained model and compared the ground truth outputs to the model's predictions. The comparison is shown in Figure 3a, where we observe a strong match (over 90% accuracy) between the predictions (blue points) and ground truths (orange points). The inset images in Figure 3a are model inputs. From this result, we can infer that the model accurately learned the key features on the optimized structures which contribute to the target wavelength-specific FOM values. Therefore, we can utilize XAI to reveal the structure-performance relationships of each device. Specifically, we employed an explanation strategy for photonics design – using SHAP – to highlight the device feature contributions to their respective

FOM[26]. These feature contribution heatmaps (represented as SHAP values, $\Phi(x,y)$, ranging from -1 to 1) are illustrated in Figure 3b, where the blue and red pixels indicate positive and negative contributions towards the FOM, respectively. We note that this is the reverse of conventional SHAP definitions due to our desired FOM being minimized. We then leverage the information captured by the XAI algorithm, and manipulate the structure accordingly, to assess its effect on device performance.

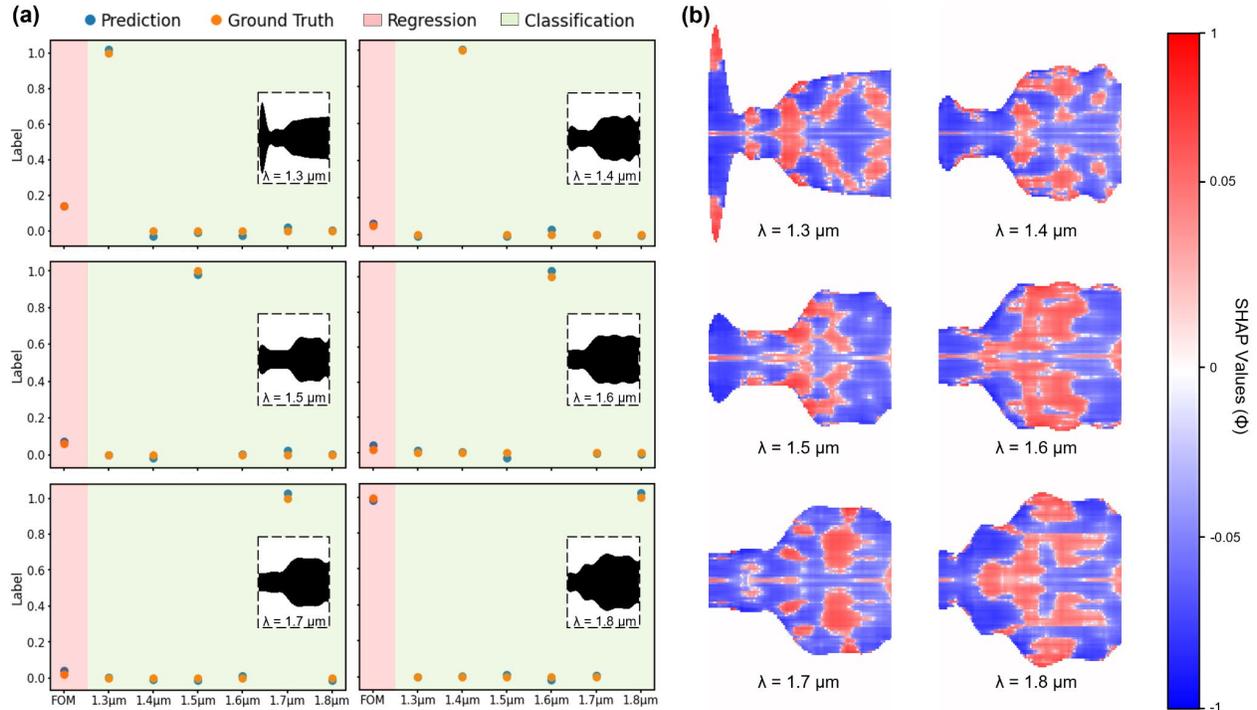

**Figure 3.** (a) Comparisons between model predictions and ground truths, for FOM (regression) and target wavelength (classification) values, show that the model accurately learned the relationship between device structure and performance. Inset images show the model inputs, which are adjoint-optimized devices. (b) SHAP explanation heatmaps of the optimized devices reveal the structural features that contribute positively (blue) or negatively (red) towards optimal device performance. Note that this is the reverse of conventional SHAP definitions due to our desired FOM being minimized.

To determine how to practically use the SHAP values (represented as red/blue heatmap pixels), we first note that high concentrations of blue pixels are located throughout a majority of each structure, while the center of the structures and select portions of the outer boundaries contain large regions of red pixels. In this regard, since the training data solely consists of geometries with varying degrees of shape changes at the SOI boundary, and no geometry change is introduced within the structure (*i.e.*, no material subtractions or white pixels are inside the main island of black pixels), we focus our analysis on the SHAP values located at the structure boundary rather than the center. Following the aforementioned principles of positive and negative contributions, we define the red regions along the structure boundaries as negative contributions towards device performance that should be removed from the design. Accordingly, we devised a boundary extraction algorithm to systematically adjust the shape of the adjoint-optimized devices using the explanation heatmaps. The algorithm, conceptually illustrated in Figure 4a, consists of an initial filtering procedure, which identifies the red-to-blue transition points along the structure boundary. In this procedure, a binarization function is applied to the SHAP values that converts the structure into existing and non-existing elements (shown in the center of Figure 4a as white and black pixels, respectively). Thresholds for binarization ($\rho(x,y)$) are given by the following step function:

$$\rho(x,y) = \begin{cases} 1 & \text{for} \quad \Phi(x,y) \leq 0, \\ 0 & \text{for} \quad \Phi(x,y) > 0, \end{cases} \tag{2}$$

where $\rho(x,y)=1$ and $\rho(x,y)=0$ indicates existing and non-existing elements, respectively. A median filter is applied to the binarization to reduce noise. Next, we "draw" a new boundary around the existing elements (Figure 4a, right) by capturing an array of points $\eta(x,y)=[X_i;Y_i]$, in which $X_i=[x_1,x_2,...,x_i]$ and $Y_i=[y_1,y_2,...,y_i]$ are vectors of length *i*. $X_i$ is an evenly spaced set of x-coordinate values from the left to the right of the image. Since *i* ultimately determines the resolution of the shape, we set its value to 20 points (matching the interval used in the initial optimization runs) to ensure that the optimizable geometry is within feasible fabrication range. This interval equates to 100 nm spacing along the x-axis, which is well within CMOS lithography resolutions. Each point on the spline can range from 0 to 1.25 µm in 20 nm steps, thus the number of parameter permutations describing the design are on the order of $1\times10^{30}$. To find the y-coordinate values in $Y_i$, we apply Algorithm 1 (detailed in the Supporting Information).

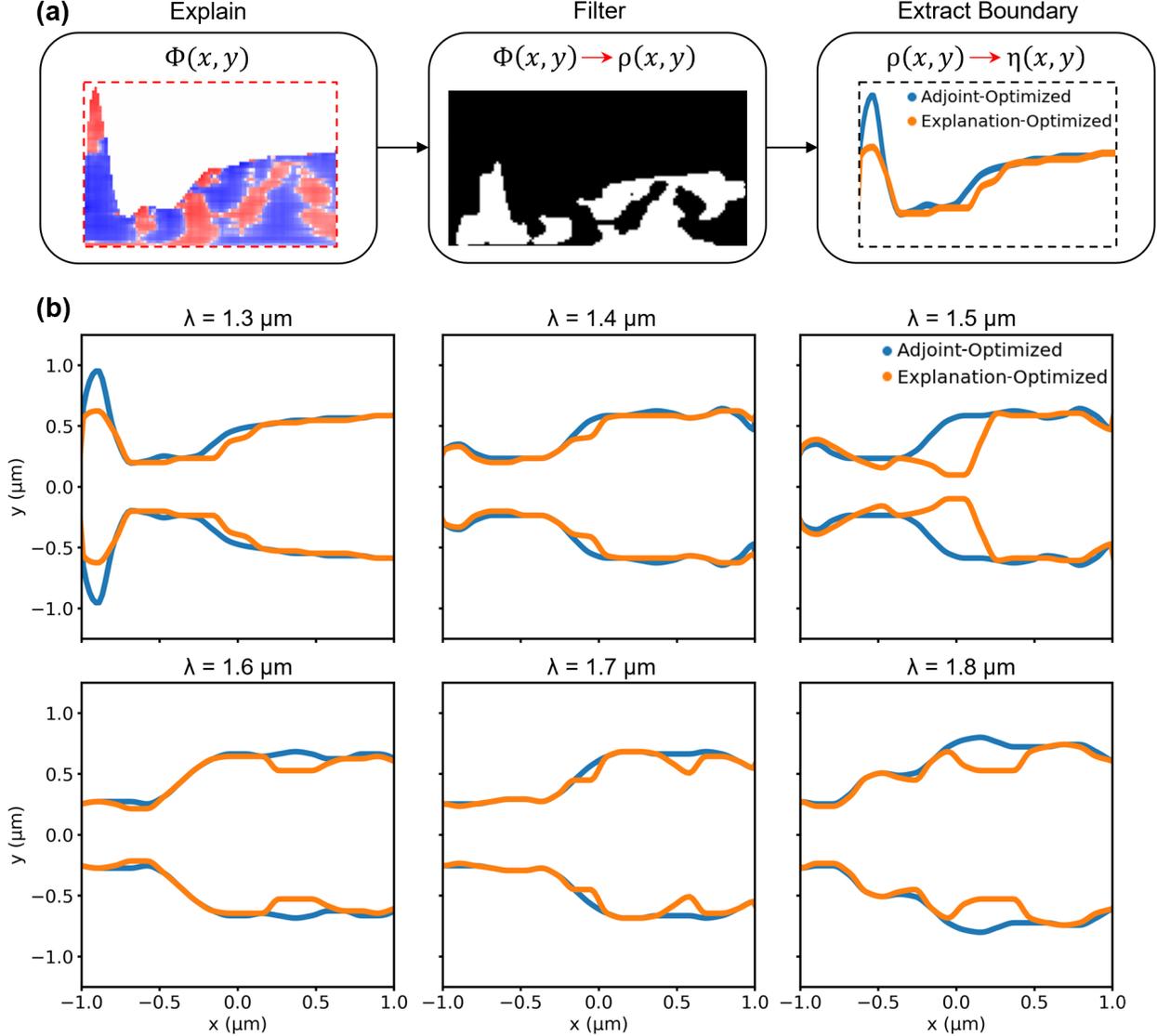

**Figure 4.** (a) Schematic representation of our explanation-optimization algorithm and workflow, consisting of explanation, filtering, and boundary extraction steps. (b) Comparison between the adjoint-optimized and explanation-optimized geometries.

Using Algorithm 1, we raster the image (from top to bottom) across all values in $X_i$ and find the points where existing elements are found (indicated by $P(x,y)=1$), then mark these points for $Y_i$. For quality purposes, we add the $\alpha$ hyperparameter to enhance robustness by reducing sharp changes in the structure as a result of filtering or noise from the explanations. We apply this workflow to each wavelength-specific adjoint-optimized structure from the previous step and present the new "explanation-optimized" results in Figure 4b. As an example of our method's application, for the 1.3 μm target design, we note that the explanation algorithm deemed the large vertical spike near the input port as a negative (red) contribution. After applying our explanation-based boundary extraction process, the height of the spike was reduced.

To assess whether the explanation (or SHAP value) based modifications to the optimized structures (*e.g.*, the spike reduction) yielded meaningful or effective contributions, we simulated the explanation-optimized designs and used them as new starting points for a second stage of adjoint optimization runs. In Figure 5, we show the FOM evolutions over the entire optimization cycle of the 35% fill fraction starting design. The explanations and optimization cycles of the remaining starting designs can be found in Figures S2-S5. The red arrows indicate the end of the first optimization stage and the beginning of the second explanation-based re-optimization stage. Further observation revealed that in the second stage of the 1.3 µm target design, reducing the vertical spike immediately reduced the FOM from 0.139 to 0.090 (a 35% improvement) at the first iteration, while the end of the optimization resulted in a final FOM of 0.050 (a 64% improvement compared to the end of the first stage). This result indicates that the explanation-based modifications overcame a saddle point in the original adjoint optimization process. In some of the other examples (*e.g.*, 1.4-1.8 µm), the first step of the second stage did not always result in an immediate FOM reduction, particularly when the FOM value was already exceedingly low (<0.075). We validate in Figure S6 that this increase in the FOM is due to the optimization getting stuck in a local minima valley instead of a saddle point, since the FOM must first increase before the algorithm can identify a lower global minima, particularly when modifying the design from where the optimization algorithm ended. However, across all the optimization targets, the end of every second-stage optimization consistently resulted in a lower final FOM than the first-stage FOM (a 39% decrease on average). Moreover, an increase in FOM followed by a further global reduction is indicative of an objective function that was previously stuck in a local minimum[14]. Thus, we demonstrate that our explanation-based re-optimization technique is capable of enhancing the performance of the adjoint optimization algorithm by allowing the FOM to escape its local minima for various optimization targets and performance ranges. We note that this entire workflow only used two optimization runs per target: one for feature contribution learning and the other for local minima escape or global FOM reduction. As previously mentioned, alternative methods at identifying lower minima typically involve repeated optimizations at random starting points or metaheuristic approaches, which can scale well-beyond two optimization runs per target (on the order of 1,000 iterations in the case of simulated annealing)[21,22].

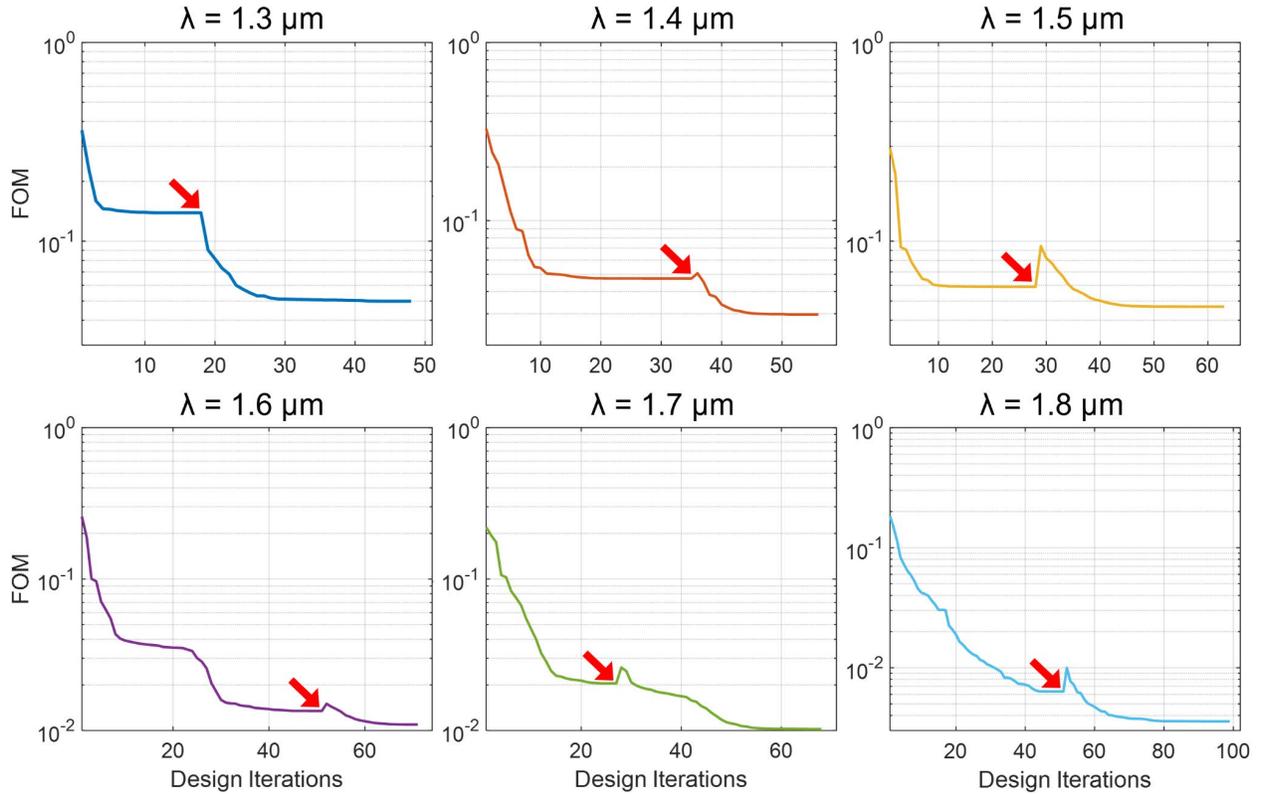

**Figure 5.** Two-stage optimization of SOI waveguide designs, across target wavelengths ranging from 1.3 μm to 1.8 μm, using the 35% fill fraction starting design. Red arrows indicate the end of the first adjoint optimization stage and the beginning of the second explanation-based re-optimization stage. Final FOM values are improved by 39%, on average, across all target wavelengths.

In the proceeding sections, we further assess the performance of the presented optimization scheme by conducting a number of additional tests, including: 1) an evaluation of the ability for SHAP to immediately improve the FOM (if the adjoint optimization is stopped prematurely) to determine the link between structure modifications and FOM improvements (*i.e.*, an "early stop" analysis), 2) the applicability of our approach on a smaller dataset and 3) different material systems, and 4) comparisons of our approach against arbitrary perturbations to the adjoint-optimized structure (*i.e.,* a "random change" analysis). First, to verify that the model is actually learning how to modify the structure, in our "early stop" analysis, we removed the portion of the training data where the adjoint optimization reached the local minima, retrained our model, and repeated our explanation-based modification. We observe in Figure S7 that across all target wavelengths, the final FOM obtained through the SHAP explanations is lower than the best available design, thereby confirming that the model is learning how to modify the structure to improve the FOM (based on the information it was given to learn).

Since there is substantial precedence in existing literature on using explainable artificial intelligence with small training datasets (on the order of several dozens to hundreds of data points), particularly to reduce the burden of data collection or computation costs[36-40], we have repeated our study on a reduced dataset using only the optimization results from a single starting design (35% fill fraction). Results of this analysis are shown in Figures S8 and S9, where we also observe performance improvements (43% on average) for all target wavelengths, which suggests that the proposed approach is applicable (to a degree) to smaller datasets.

Next, we evaluated the generalizability of the proposed framework by applying it to other contemporary nanophotonics design challenges. We note that prior studies have successfully applied XAI to alternative nanophotonic structures, such as metasurfaces, and demonstrated performance enhancements in the form of spectral property tuning[26] (though no optimization algorithm integration was employed). Thus, we focused this generalizability analysis on material alternatives. In this regard, over the past few years, $Si_3N_4$ has emerged as a promising alternative to silicon in photonic systems. Compared to silicon, $Si_3N_4$ has lower propagation losses and does not exhibit two-photon absorption in the telecommunications range[31-34] (among other pros and cons). Accordingly, we performed the same two-stage optimization study on a $Si_3N_4$ waveguide (for the same Y-splitter starting geometry) and found that the second-round optimizations were also able to surpass the results of the first at every target wavelength (Figure S10). Across all the test cases, an average FOM improvement of 74% was achieved.

Lastly, to show that the achieved performance enhancements were not simply obtained through arbitrary perturbations to the optimized structure, we performed an additional "random change" analysis where we randomly modified the first-stage structures, repeated the second stage of optimizations, and compared the results. This comparison is presented in Figure S11, where five random modifications (defined in the Supporting Information) were made to each structure. Across the 30 tests performed on the six optimized designs, all of the randomly modified structures possess higher FOM values (*i.e.,* lower performance) than those of the explainability-optimized devices, while only two "random change" results fall within 25% of the explainability-optimized device performance. Additionally, 28 tests produced higher final FOM values than the initial optimized designs. Therefore, not only are the random changes ineffective in terms of escaping the local minima, but they can also inadvertently push the optimization into a worse state than where the optimization started at. As such, we demonstrate that our XAI-based approach is not stochastic in nature, but can deterministically tune a structure in order to maximize performance. Through the preceding series of tests, we show that the presented approach is generally applicable to numerous applications of adjoint optimization for electromagnetic design, including those with different constituent materials, structures, and optimization targets.

## Conclusions

In summary, we present an inverse design framework that extends the capabilities of gradient-based shape or topology optimization algorithms for photonic inverse design, while elucidating the relationships between device performance and nanoscale structuring. Our framework combines adjoint optimization, AutoML, and XAI to enhance device performance beyond that which is obtainable through the optimization algorithm alone. We applied our method to SOI waveguide design and showed that the optimization algorithm initially reaches a performance plateau (*i.e.*, local minima). After utilizing XAI to reveal the device's structural contributions towards a designated FOM (where 0 represents ideal performance), we leveraged this information (in conjunction with a boundary extraction algorithm) to push the optimization out of its local minima and reduce the FOM further. Across a range of performance-plateaued devices optimized for various wavelengths within the 1.3 μm to 1.8 μm telecom window, our method was able to improve device performance by an average of 39%. The entire procedure only requires two optimization runs per optimization target, which is potentially more computationally efficient than alternative approaches, particularly those that rely on multiple optimization runs and random starting points. Additionally, generalizability tests performed on $Si_3N_4$ waveguides showed an average of 74% device performance improvement. Thus, we demonstrate that our XAI-based approach provides an automated and systematic solution for an electromagnetic optimization algorithm to escape local minima and achieve greater device performance. Looking ahead, integrating conventional optimization and data-driven machine learning will likely prove a fruitful direction for inverse design and physics discovery in photonic systems.

# Supporting Information

AutoML training results, details of the boundary extraction algorithm, random starting design analysis, local minimum analysis, early stop analysis, reduced dataset analysis, random change comparison, generalizability or material alternative assessment, and overfitting analyses.

# Acknowledgements

This work used computational and storage services associated with the Hoffman2 Shared Cluster provided by UCLA Institute for Digital Research and Education's Research Technology Group.

# Funding Sources

This work is supported by the Sloan Research Fellowship from the Alfred P. Sloan Foundation and the DARPA Young Faculty Award (#W911NF2110345).

# Conflicts of interest

The authors declare no conflicts of interest.

# Supporting Information

**AutoML Training and Execution**

In this work, we use neural architecture search (AutoKeras) with image blocks as well as an early stopping callback to automate the deep learning process by testing different model variants across multiple trials. We observe that the optimal architecture was identified after 12 trials, which has a validation loss of $9.1 \times 10^{-5}$. The final training and validation losses of each trial are presented in Figure S1a, and the evolution of the convolutional neural network (CNN) architecture from the first trial to the last is shown in Figure S1b. The first trial initially begins with a two-block architecture and a validation loss of $3.8 \times 10^{-2}$. At the final trial, the optimized CNN possesses five convolutional blocks (with 512, 256, 128, 64, and 32 filters, respectively) followed by a dense layer. Each block contains Leaky ReLu, batch normalization, and max pooling layers. The final model possesses approximately 1.5 million training parameters. We note that the number of parameters are attributed to the size of the input images (128×64 pixels).

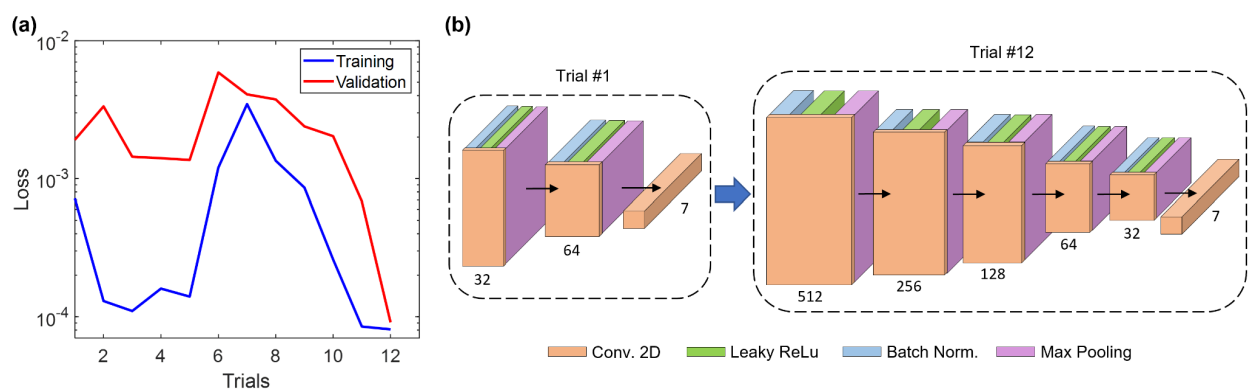

**Figure S1.** (a) Training and validation loss progression across the AutoML trials. (b) Schematic of the model architecture changes between the first and final trials.

**Boundary Extraction Algorithm**

As described in the main text, we devised a boundary extraction algorithm to systematically adjust the shape of the adjoint-optimized devices using the explanation heatmaps. This algorithm is represented as:

**Algorithm 1**: Boundary Extraction

---

**Input:** Filtered image array ($P$) of $\rho(x,y)$ values at every image pixel.

**for** $x \in X_i$ **do**

   **for** $y = 1\!:\!L$, where $1$ is the top of the image and $L$ is the height or bottom of the image, **do**

      **if** $P(x,y) = 1$ and $y_i - y_{i-1} < \alpha$, where $\alpha$ is a hyperparameter for enhancing robustness, **then**

         $Y_i.append(y_i)$

      **else**

         $Y_i.append(y_{i-1})$

**Return:** $Y_i$

---

As shown in Algorithm 1, we raster the image (from top to bottom) across all values in $X_i$ and find the points where existing elements are found (indicated by $P(x,y)=1$), then mark these points for $Y_i$. For quality purposes, we add the $\alpha$ hyperparameter (with a default value of 5 points or 100 nm in device length) to enhance robustness by reducing sharp changes in the structure as a result of filtering or noise from the explanations.

**Random Starting Designs and Explanation Results**

    We generated the data for our model by performing adjoint optimization runs on randomized starting designs. Figure S2 below shows all the starting designs, where S2a, S2b, S2c, and S2d (first column) correspond to 25%, 35%, 40%, and 50% fill fractions, respectively. The optimization runs on the presented starting designs (at the designated target wavelengths) yielded a training dataset of over 600 input-output pairs. After training our model on this larger dataset, SHAP explanations were captured for each optimized structure (second column of Figure S2) and re-optimized following the same procedure described in the main text (third column of Figure S2).

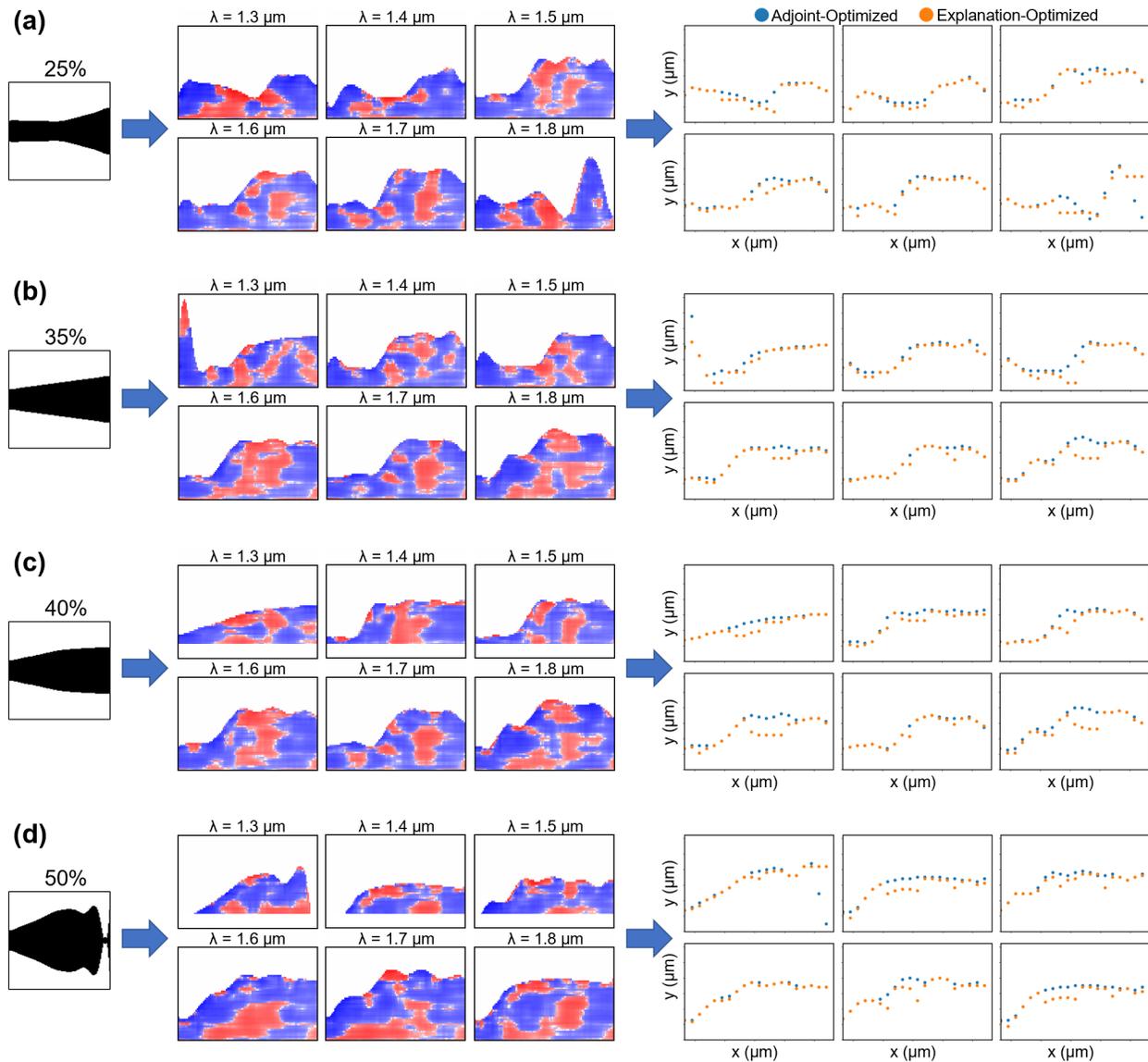

**Figure S2.** Randomized adjoint optimization starting designs with (a) 25%, (b) 35%, (c) 40%, and (d) 50% fill fractions. Device layouts (first column) and their corresponding target wavelength-specific adjoint-optimized device explanations (second column) are illustrated. Top halves of the structures are presented for ease of visualization. Explanations are used to identify new starting points for optimization, which can escape local minima (third column). Optimization results of the presented designs are shown in Figures 5 and S3-S5.

In Figures S3, 5, S4, and S5, we present the optimization cycles of the 25%, 35%, 40%, and 50% fill fraction starting designs, respectively. The red arrow indicates the end of the first optimization stage and the beginning of the second explanation-based re-optimization stage. Across each starting design, the end of every second-stage optimization consistently resulted in a lower final FOM than the first-stage FOM. Specifically, the 25% fill fraction starting design ended with a 47% average improvement across all target wavelengths, while the 35%, 40%, and 50% fill fractions resulted in 39%, 45%, and 45% improvements, respectively.

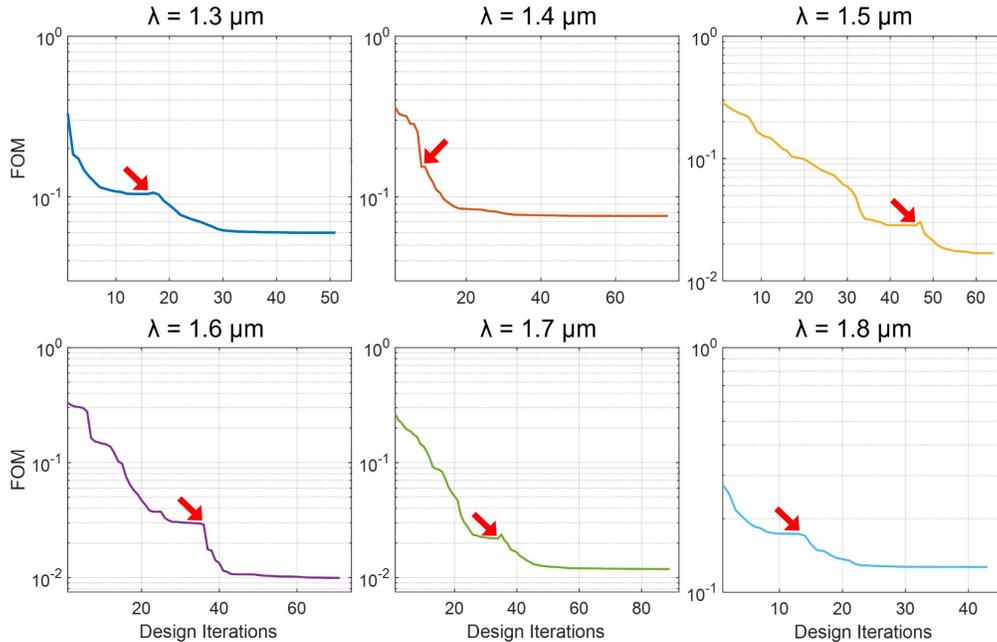

**Figure S3.** Two-stage optimization of SOI waveguide designs, across target wavelengths ranging from 1.3 μm to 1.8 μm, using the 25% fill fraction starting design shown from Figure S2. Red arrows indicate the end of the first adjoint optimization stage and the beginning of the second explanation-based re-optimization stage. Final FOM values are improved by 47%, on average, across all target wavelengths.

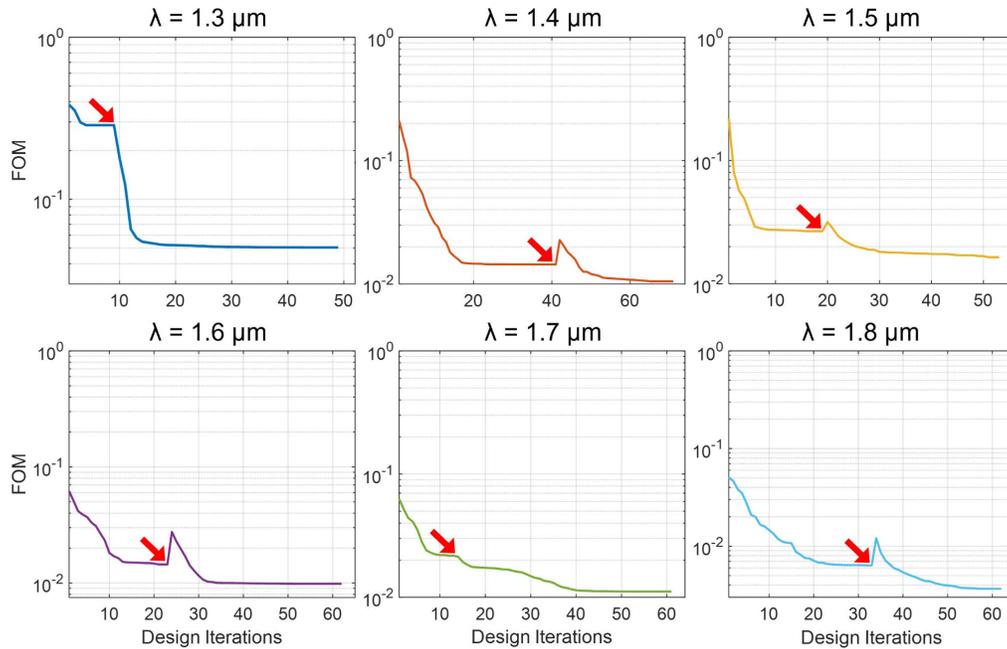

**Figure S4.** Two-stage optimization of SOI waveguide designs, across target wavelengths ranging from 1.3 μm to 1.8 μm, using the 40% fill fraction starting design shown from Figure S2. Red arrows indicate the end of the first adjoint optimization stage and the beginning of the second explanation-based re-optimization stage. Final FOM values are improved by 45%, on average, across all target wavelengths.

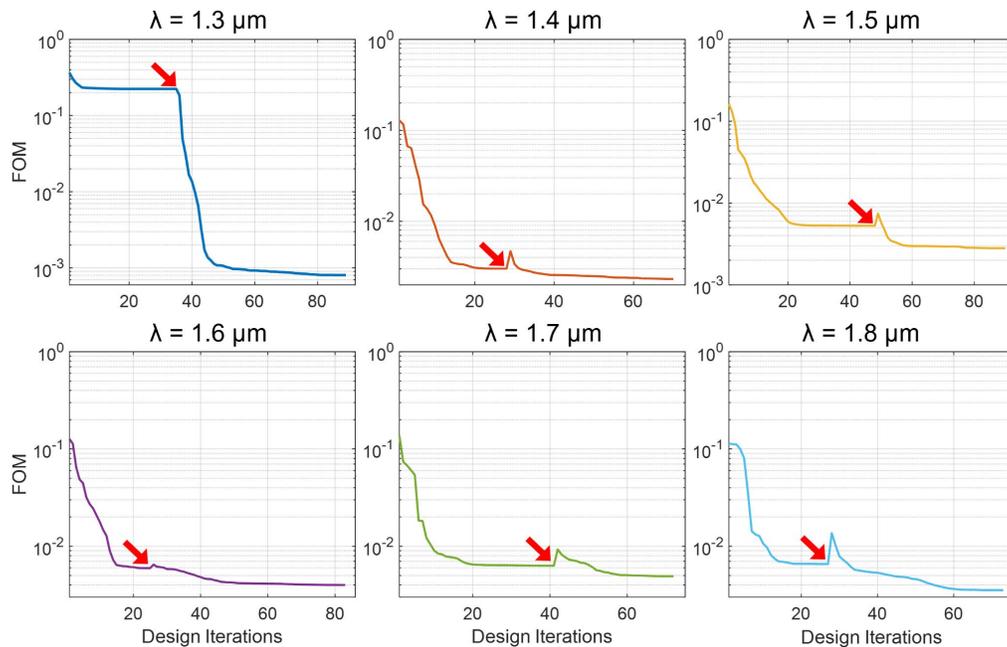

**Figure S5.** Two-stage optimization of SOI waveguide designs, across target wavelengths ranging from 1.3 μm to 1.8 μm, using the 50% fill fraction starting design shown from Figure S2. Red arrows indicate the end of the first adjoint optimization stage and the beginning of the second

explanation-based re-optimization stage. Final FOM values are improved by 45%, on average, across all target wavelengths.

**Local Minima Analysis**

We demonstrate that the adjoint-optimized designs are trapped in a local minimum by attempting to improve a corresponding design further through exhaustive exploration. Here, we focus our analysis on the waveguide structure from the main text (Figure 3) that is optimized for power transmission at 1.8 μm. Using this structure, we individually modify each of the 20 parameters that make up the overall shape of the waveguide (conceptually illustrated in Figure S6a). In Figure S6b, we observe that each of these changes (40 new simulations total) yield a higher FOM than the original adjoint-optimized design. Thus, the design here has reached a local minimum and cannot be improved through gradient-based optimization alone. However, since the adjoint-optimized design was obtained after more than 50 iterations, an alternative optimization route may be identified by utilizing an earlier design iteration as the base of the analysis, but this would require a large number of additional simulations. Furthermore, if multiple interdependent parameters or points are considered, this can exponentially increase computation time and requirements. Therefore, the presented XAI approach simply augments this process by rapidly guiding the optimization towards a location where the adjoint method or sensitivity analysis performs better. This is indicated by the dashed lines in Figure S6b, where we show that the FOM of the explanation-optimized design is lower than that obtained through adjoint optimization.

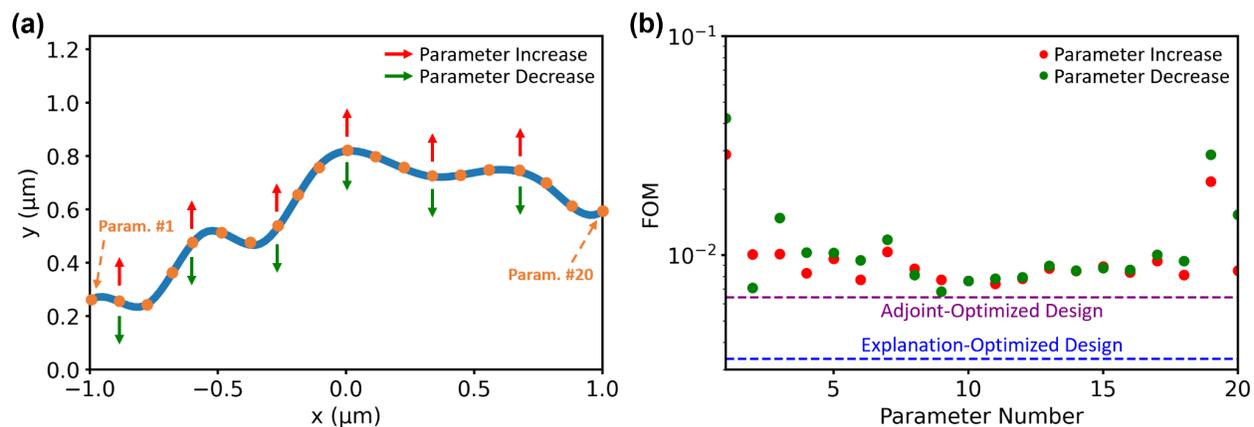

**Figure S6.** Local minimum analysis of an adjoint-optimized design. (a) Schematic illustration of the analysis approach, where each shape parameter is individually modified by either increasing (red) or decreasing (green) the size or spatial coordinate of the structure by 50 nm, while all other parameters remain fixed. (b) The resulting FOM values of the corresponding parameter changes

show that the adjoint-optimized design is stuck in a local minimum, while the explanation-optimized design can improve the design further by finding a new optimization route.

**Early Stop Analysis**

To verify that the model is actually learning how to modify the structure, we evaluate the ability for SHAP to immediately improve the FOM to determine the link between structure modifications and FOM improvements. Since in the main text, the SHAP modifications either yield an increased or decreased FOM (due to the presence of either a saddle point or local minimum valley), this link is not apparent. To perform this analysis, we removed the portion of the training data where the adjoint optimization reached the local minima, retrained our model, and repeated the explanation-based modification. In this "early stop" analysis (shown in Figure S7), the solid lines represent the optimization data that is included in the training dataset, and the (*) marker represents the FOM of the modified structure. We note that across all target wavelengths, the final FOM obtained through the SHAP explanations is lower than the best available design, thereby confirming that the model is learning how to modify the structure to improve the FOM (based on the information it was given to learn).

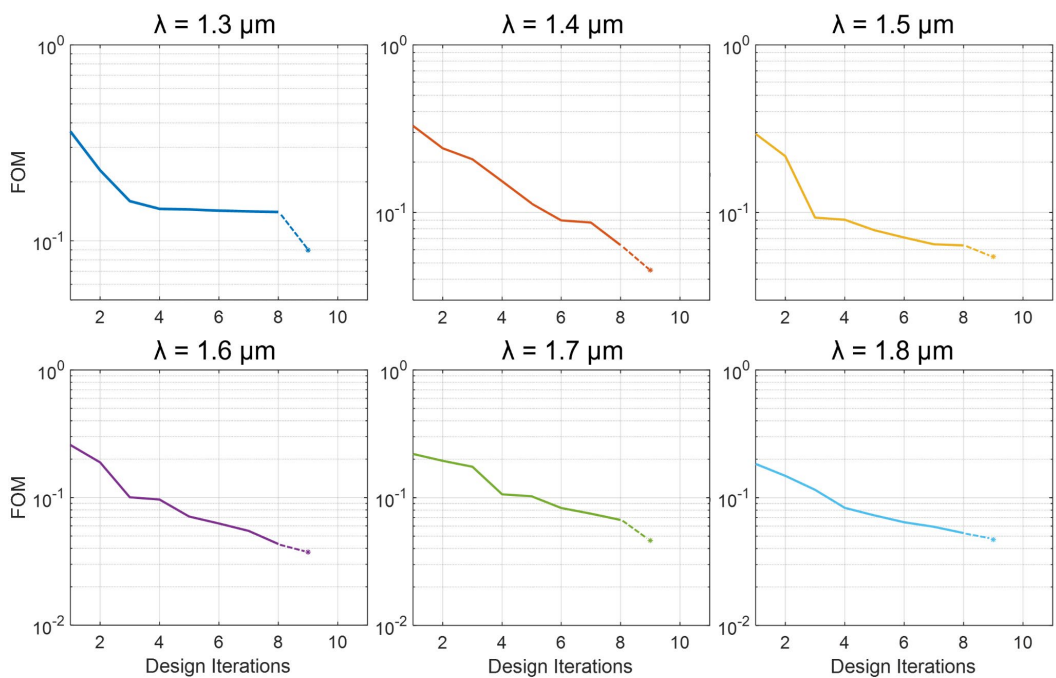

**Figure S7.** Explanation-modified structures, indicated by the dashed lines leading to the "*" markers, when the adjoint optimization is stopped prematurely. Modified structures possess lower

FOM values than the best structure from the training dataset, which further indicate that explanation-based modifications are linked to FOM improvements.

**Reduced Dataset Analysis**

Since there is substantial precedence in existing literature on using explainable artificial intelligence with small training datasets (on the order of several dozens to hundreds of data points), particularly to reduce the burden of data collection or computation costs, we have repeated our study on a reduced dataset using only the optimization results from a single starting design (35% fill fraction), which yielded 192 input-output pairs. Using this reduced dataset, AutoML identified the optimal model architecture after 17 trials, which has a validation loss of $8.6\times10^{-5}$. Similar to the main text, Figure S8 compares the model predictions (blue points) with the ground truths (orange points) using this reduced dataset. From this result, we can infer that the model accurately learned the key features on the optimized structures which contribute to the target wavelength-specific FOM values. Therefore, we can utilize XAI to reveal the structure-performance relationships of each device.

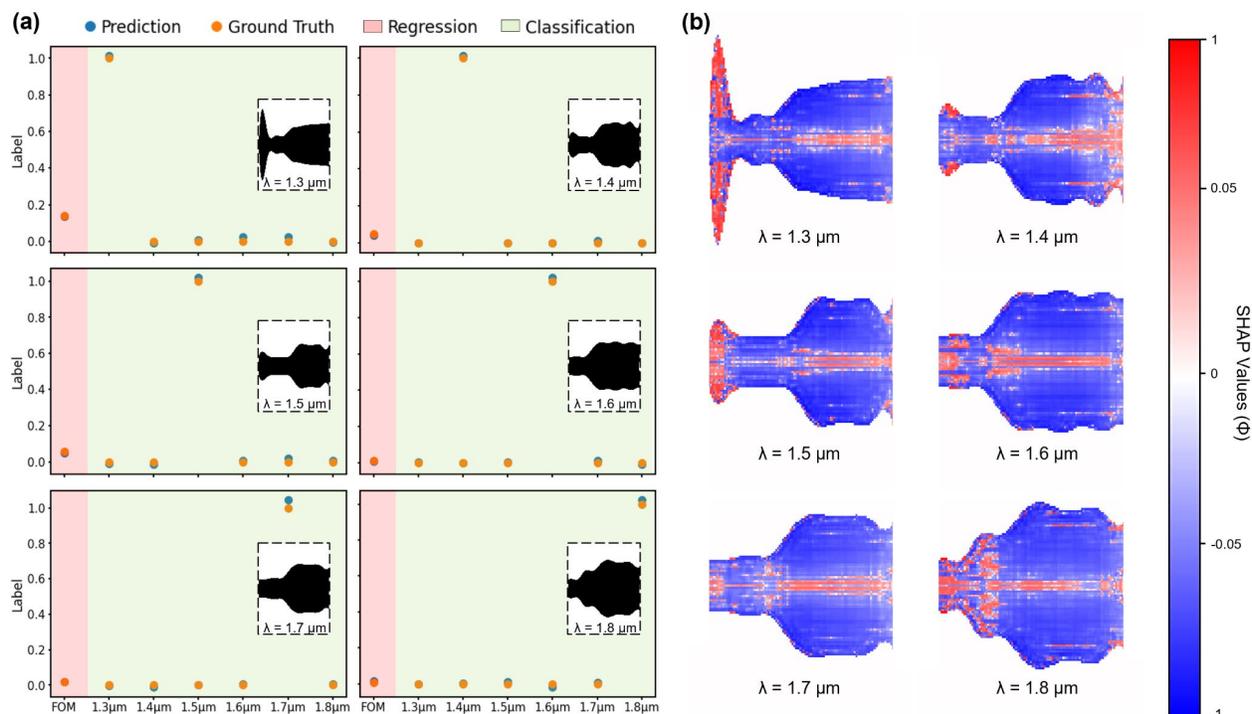

**Figure S8.** (a) Comparisons between model predictions and ground truths, for FOM (regression) and target wavelength (classification) values, show that the model accurately learned the relationship between device structure and performance. Inset images show the model inputs, which are adjoint-optimized devices. (b) SHAP explanation heatmaps of the optimized devices reveal the structural features that contribute positively (blue) or negatively (red) towards optimal device performance.

In Figure S9, we show the FOM evolutions over the entire optimization cycle of the 35% fill fraction starting design, and similarly observe that the end of every second-stage optimization consistently resulted in a lower final FOM than the first-stage FOM (by 43% on average). We note that the reduced dataset (with only the 35% fill fraction starting design) performed slightly better than the larger-data model for this particular starting design (a 39% improvement for the larger dataset in comparison to a 43% improvement for the reduced dataset). Thus, we demonstrate that the presented approach can yield performance improvements in a wide range of randomized starting points. However, the degree of performance improvement may be linked to the size and breadth of the training data, which we aim to investigate in future works.

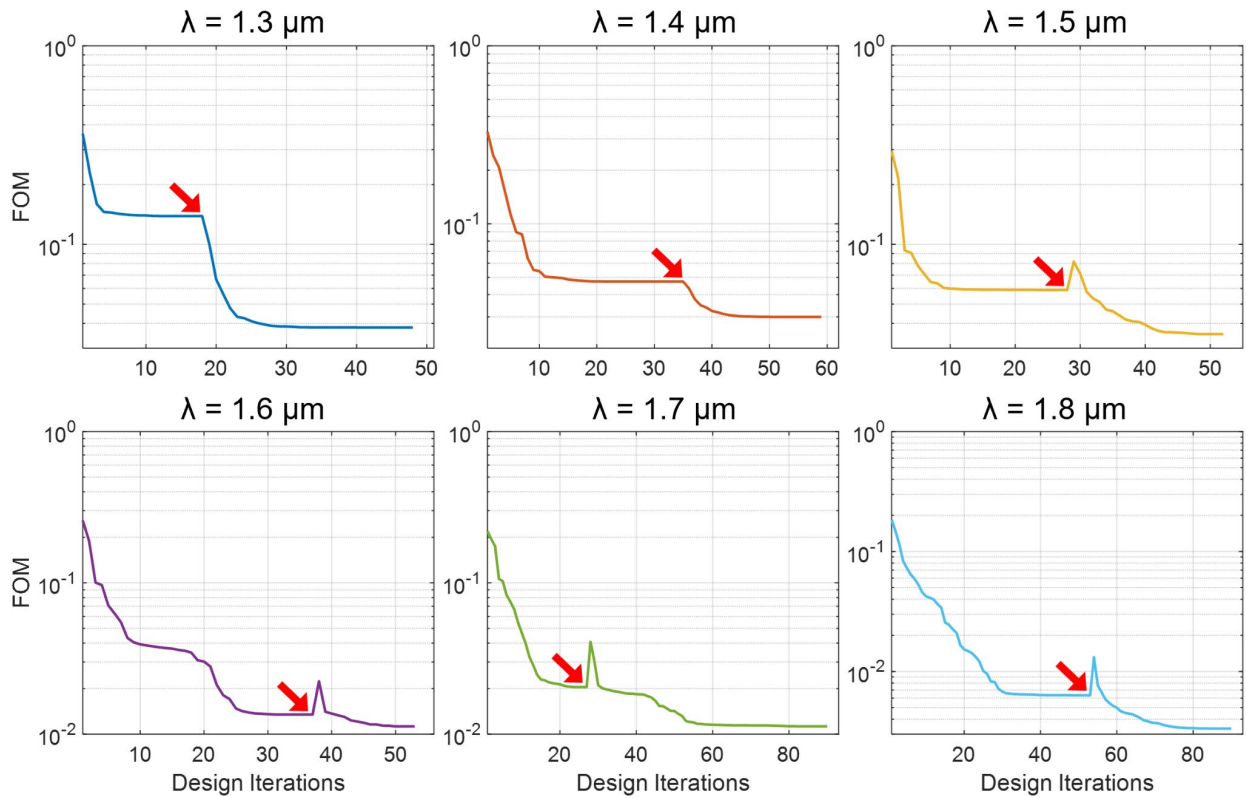

**Figure S9.** Two-stage optimization of SOI waveguide designs across target wavelengths ranging from 1.3 μm to 1.8 μm. Red arrows indicate the end of the first adjoint optimization stage and the beginning of the second explanation-based re-optimization stage. Final FOM values are improved by 43%, on average, across all target wavelengths.

**Random Change Comparison**

To show that the achieved performance enhancements were not simply obtained through arbitrary perturbations to the optimized structure, we performed an additional "random change" analysis where we randomly modified the first-stage structures, repeated the second stage of optimizations, and compared the results. This comparison is presented in Figure S10, where five random modifications were made to each structure. Random modifications were introduced by adding 300 nm to $y_n$ and $y_{n+1}$ in the optimizable parameter vector $Y_i$, where $n$ is varied from 1 to $i-1$ (where $i$ is the length of the vector). Across the 30 tests performed on the six optimized designs, all of the randomly modified structures possess higher FOM values (*i.e.,* lower performance) than the explainability-optimized devices, while only two results fall within 25% of the latter's final device performance. Additionally, 28 tests yield higher final FOM values than the initial optimized design. Therefore, not only are the random changes ineffective in terms of escaping the local minima, but they can also inadvertently push the optimization into a worse state than where the optimization started at.

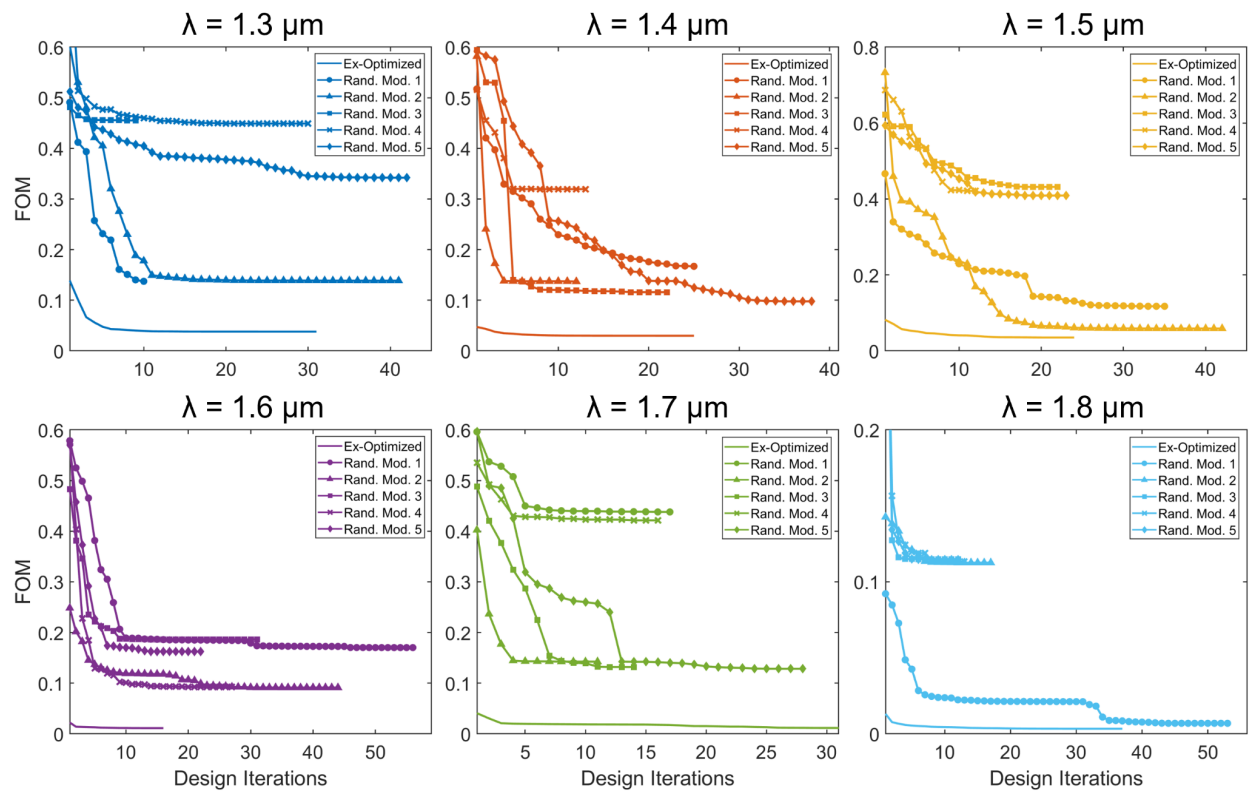

**Figure S10.** Comparison between the explanation-optimized and randomly modified structures. Random modifications either produce FOM values that are higher (lower performance) than the explanation-optimized structures, or higher than the starting point of optimization.

**Generalizability or Material Alternative Assessment**

We evaluated the generalizability of the proposed framework by applying it to other contemporary nanophotonics design challenges. In this regard, we performed the same two-stage optimization study on a $Si_3N_4$ waveguide (for the same Y-splitter starting geometry). Similar to the results presented in the main text (Figure 5), here we observe that the second stage (red-boxed region) of each target wavelength-specific optimization produces a lower overall FOM than the final FOM obtained through adjoint optimization alone (indicated by the red arrows). In particular, as shown in Figure S11, we observe a minimum of 31% FOM improvement (for the 1.6 µm design) and a maximum of 90% improvement (for the 1.3 µm design), for an average of 74% improvement across all the test cases. As a result, we show that the presented approach is generally applicable to numerous applications of adjoint optimization for electromagnetic design, including those with different constituent materials, structures, and optimization targets.

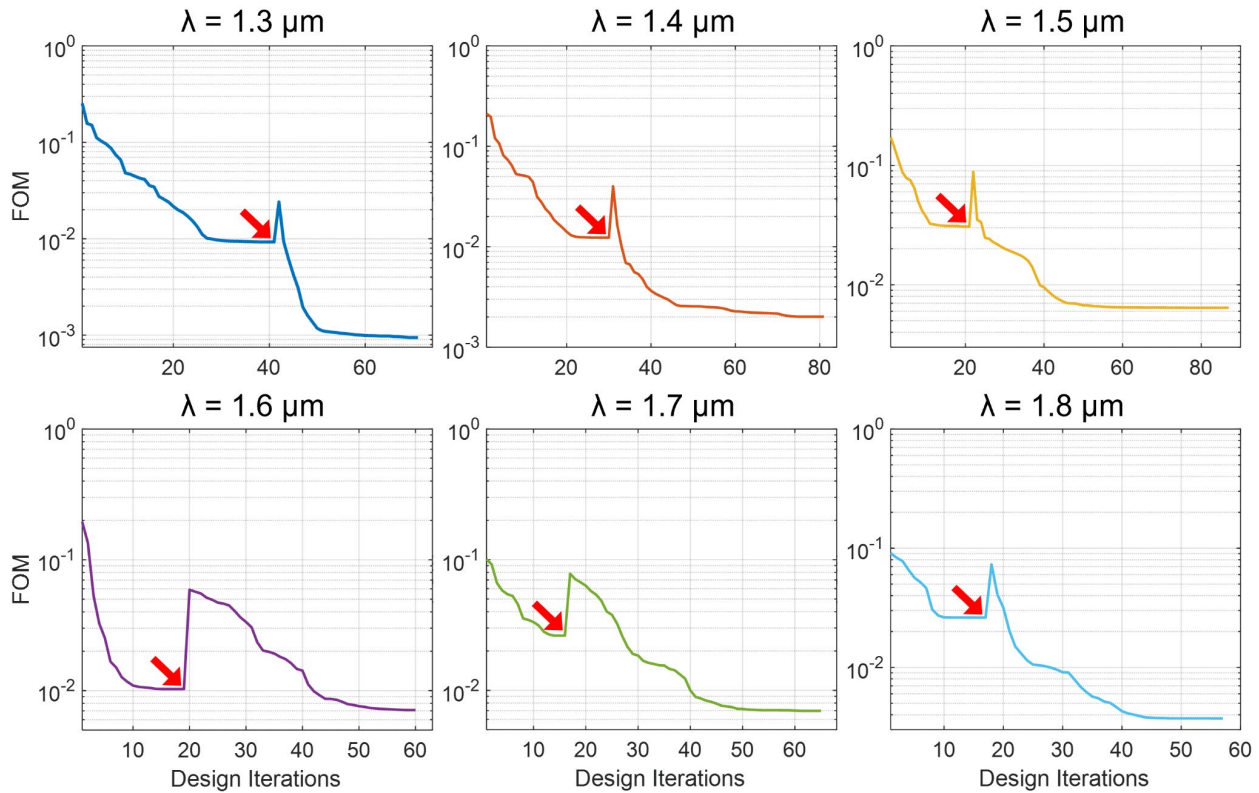

**Figure S11.** Two-stage optimization of Si$_3$N$_4$ waveguide designs across target wavelengths ranging from 1.3 μm to 1.8 μm. Red arrows indicate the end of the first adjoint optimization stage and the beginning of the second explanation-based re-optimization stage. Final FOM values are improved by 74%, on average, across all target wavelengths.

**Cross Validation and Overfitting Analysis**

To ensure that our model did overfit during or after training, we performed a number of additional tests. First, we performed k-fold cross validation. Here, the dataset was split into 'k' consecutive folds, and each fold was used once as validation while the remaining 'k-1' folds formed the training data (at each fold number), thus the entire dataset was used to validate our model. For a more thorough analysis, we performed 10-fold and 5-fold cross validations on our optimized model, and presented the results in Tables S1 and S2, respectively. 10-fold cross validation (where 90% of the data was used for training and 10% for validation, at each fold, number after shuffling the data) resulted in an average MSE of $2.1\times10^{-4}$ with a standard deviation of 0.01%. On the other hand, 5-fold cross validation (where 80% of the data was used for training and 20% for validation, at each fold, number after shuffling the data) resulted in an average MSE of $2.9\times10^{-4}$ with a standard deviation of 0.01%. Since the MSE values derived from the 5-fold validation were considerably larger than those obtained from the 10-fold validation, it can be inferred that an increased amount of training data (90% vs. 80%) contributed to increased model performance. Additionally, across both cross validation procedures, the validation losses were consistently low across each fold (evident from the 0.01% standard deviation between folds), which indicates that our model is not overfitting against a particular validation dataset.

**Table S1.** K-fold cross validation results (k=10).

| Fold Number | Validation Loss (MSE) |
|---|---|
| 1 | $1.2019\times10^{-4}$ |
| 2 | $3.0729\times10^{-4}$ |
| 3 | $3.7655\times10^{-4}$ |
| 4 | $9.0292\times10^{-5}$ |
| 5 | $2.3388\times10^{-4}$ |
| 6 | $1.4062\times10^{-4}$ |
| 7 | $1.5254\times10^{-4}$ |
| 8 | $3.8326\times10^{-4}$ |
| 9 | $1.9175\times10^{-4}$ |
| 10 | $1.3702\times10^{-4}$ |
| **Average MSE** | $2.1334\times10^{-4}$ |

| Standard Deviation (%) | 0.010742 |

Table S2. K-fold cross validation results (k=5).

| Fold Number | Validation Loss (MSE) |
|---|---|
| 1 | $2.0566 \times 10^{-4}$ |
| 2 | $1.8427 \times 10^{-4}$ |
| 3 | $2.5280 \times 10^{-4}$ |
| 4 | $4.1815 \times 10^{-5}$ |
| 5 | $4.0167 \times 10^{-4}$ |
| **Average MSE** | $2.9251 \times 10^{-4}$ |
| **Standard Deviation (%)** | 0.0110 |

For another detailed look at model performance, we performed a goodness of fit analysis on our model and included a new test set that was derived from the previous training and validation datasets. In this new analysis, we used 80% of the data for training, 10% for validation, and 10% for testing. To adhere to standard machine learning practices, we note that this test set was used to evaluate the model *after* training, whereas the validation set was used to evaluate the model *during* training. Losses for the training, validation, and test sets are $2.0 \times 10^{-4}$, $2.5 \times 10^{-4}$, and $2.6 \times 10^{-4}$, respectively. We believe this result shows that the test set, previously not seen by the neural network in either training or validation, has almost exactly the same loss value as the validation set, which is a strong indication of generalization. Additionally, Figure S12 shows the ground truth and prediction comparisons for each datapoint, where we observe that our model's predictions are accurate for both high and low performance designs across each dataset (training, validation and test), thus further indicating that the model is not overfitting or skewed towards near-optimal devices.

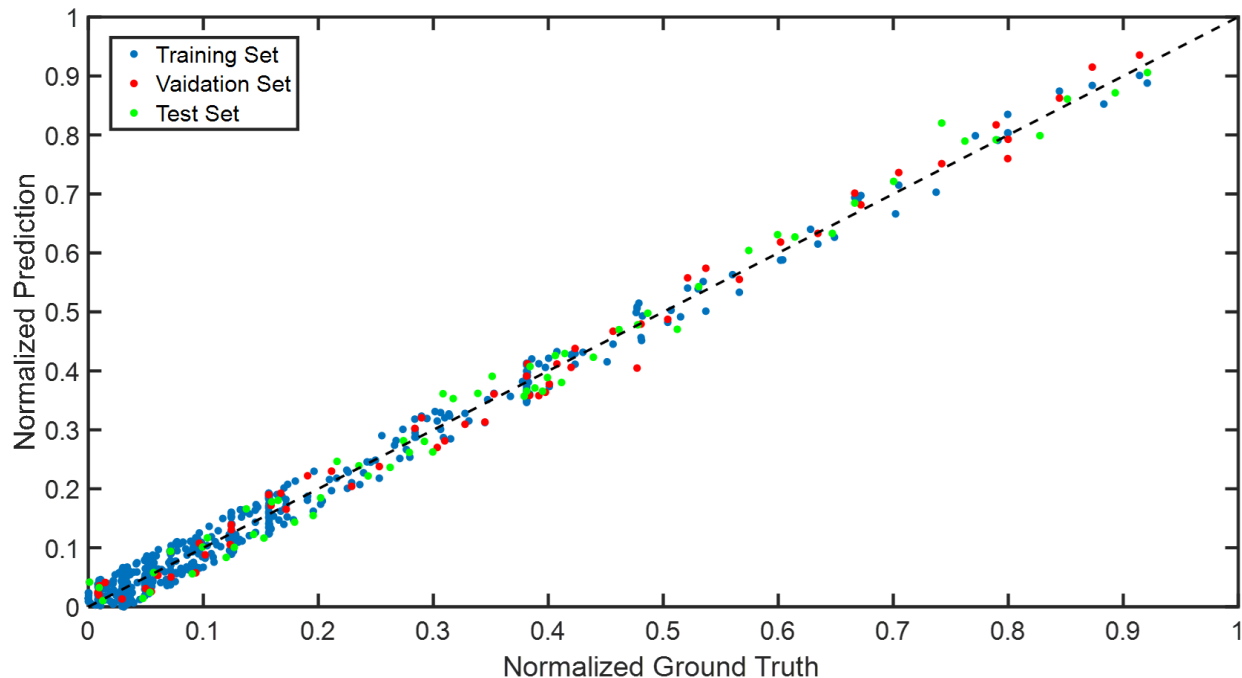

**Figure S12.** Goodness of fit analysis on model predictions vs. ground truth (normalized) in the training (80%), validation (10%), and test (10%) sets. Losses for the training, validation, and test sets are $2.0\times10^{-4}$, $2.5\times10^{-4}$, and $2.6\times10^{-4}$, respectively.

In the original training process, we included an early stopping callback function to ensure that training automatically terminates once the validation loss shows no further improvement. However, to determine whether our model will eventually overfit during training (such that we can confirm that our model did not overfit over the allotted epochs), we removed the early stop callback and retrained the model up to 1000 epochs. In Figure S13 below, we observe that the training and validation losses begin to diverge at approximately 400 epochs (where the training was previously terminated), which indicates that the model is memorizing the training data after 400 epochs.

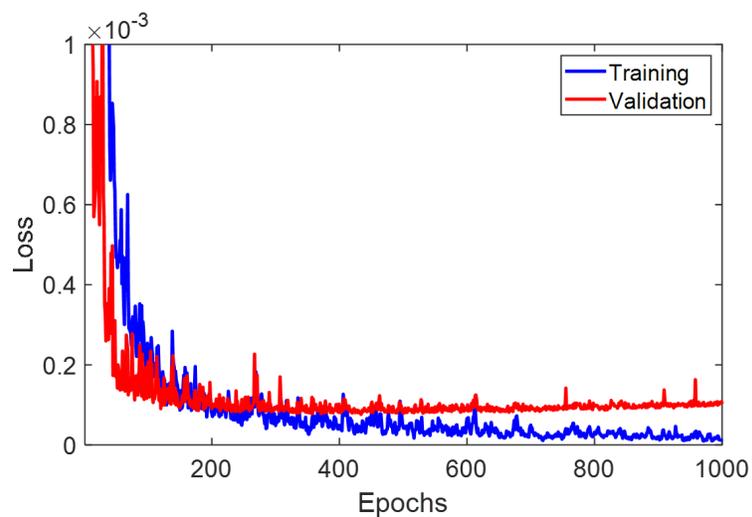

**Figure S13.** Model training without early stopping. Training and validation losses begin to diverge at approximately 400 epochs, thus indicating that the final model did not overfit the training data.